\documentclass[journal=langmuir,manuscript=article,amsmath]{achemso}
\usepackage[version=3]{mhchem} 

\author{D. A. Matoz-Fernandez}
\author{D. H. Linares}
\author{A. J. Ramirez-Pastor}
\email{antorami@unsl.edu.ar} \affiliation[UNSL] {Departamento de
F\'{\i}sica, Instituto de F\'{\i}sica Aplicada, Universidad
Nacional de San Luis-CONICET, Chacabuco 917, D5700BWS San Luis,
Argentina}

\title{Statistical thermodynamics of long straight rigid rods on triangular lattices: nematic order and adsorption thermodynamic functions}

\begin{document}

\begin{abstract}
The statistical thermodynamics of straight rigid rods of length
$k$ on triangular lattices was developed on a generalization in
the spirit of the lattice-gas model and the classical
Guggenheim-DiMarzio approximation. In this scheme, the Helmholtz
free energy and its derivatives were written in terms of the order
parameter $\delta$, which characterizes the nematic phase
occurring in the system at intermediate densities. Then, using the
principle of minimum free energy with $\delta$ as a parameter, the
main adsorption properties were calculated. Comparisons with Monte
Carlo simulations and experimental data were performed in order to
evaluate the reaches and limitations of the theoretical model.
\end{abstract}

\section{1. Introduction}

The adsorption of gases on solid surfaces is a topic of
fundamental interest for various applications \cite{Critte,Keller}. From the theoretical point
of view, the process can be described in terms of the lattice-gas
model \cite{Steele,Dash,Dash1,Shina,Binder,Patrykiejew}. A lattice gas is a system
of $N$ molecules bound not more than one per site to a set of $M$
equivalent, distinguishable, and independent sites, and without
interactions between bound molecules. Many studies have been
carried out on the adsorption behavior of small molecules in such
systems. However, the problem in which a 2D lattice contains
isolated points (vacancies) as well as $k$-mers (particles
occupying $k$ adjacent sites) has not been solved in closed form
and still represents a major challenge in surface science.

A previous paper \cite{LANG14} was devoted to the
study of long straight rigid rods adsorbed on square lattices. In
Ref. [\cite{LANG14}], the Helmholtz free energy of the
system and its derivatives were written in terms of the order
parameter $\delta$, which characterizes the nematic phase
occurring in the system at intermediate densities \cite{Ghosh,JCP7}. Then, using the principle of
minimum free energy with $\delta$ as a parameter, the main
adsorption properties were calculated. Comparisons with Monte
Carlo (MC) simulations revealed that the new thermodynamic
description was significantly better than the existing theoretical
models developed to treat the polymer adsorption problem.

In contrast to the statistic for the simple particles, where the
arrangement of the adsorption sites in space is immaterial, the
structure of lattice space plays such a fundamental role in
determining the statistics of $k$-mers. Then, it is of interest
and of value to inquire how a specific lattice structure
influences the main thermodynamic properties of adsorbed
polyatomics. In this sense, the aim of the present work is to
extent the study in Ref. [\cite{LANG14}] to
triangular lattices. The problem is not only of theoretical
interest, but also has practical importance. A complete summary
about adsorption on triangular lattices can be found in \cite{Phares1,Phares2,Phares3,Phares4} and references therein.

The rest of the paper is organized as follows. In Section 2, the
theoretical formalism is presented. Section 3 is devoted to
describe the Monte Carlo simulation scheme.  The analysis of the
results and discussion are given in Section 4. Finally, the
conclusions are drawn in Section 5.

\section{2. Model and theory}

In this paper, the adsorption of straight rigid rods (or $k$-mers)
on triangular lattices is considered. The adsorbate molecules are
assumed to be composed by $k$ identical units in a linear array
with constant bond length equal to the lattice constant $a$. The
$k$-mers can only adsorb flat on the surface occupying $k$ lattice
sites. The substrate is represented by a triangular lattice of $M
= L \times L$ adsorption sites, with periodic boundary conditions.
$N$ particles are adsorbed on the substrate with 3 possible
orientations along the principal axis of the array [see $k$-mers
marked with 1, 2 and 3 in Figure 1(a)]. The only interaction between
different rods is hardcore exclusion: no site can be occupied by
more than one $k$-mer unit. The surface coverage (or density) is
defined as $\theta=kN/M$.

\begin{figure*}
\begin{center}
\includegraphics[width=0.85\textwidth]{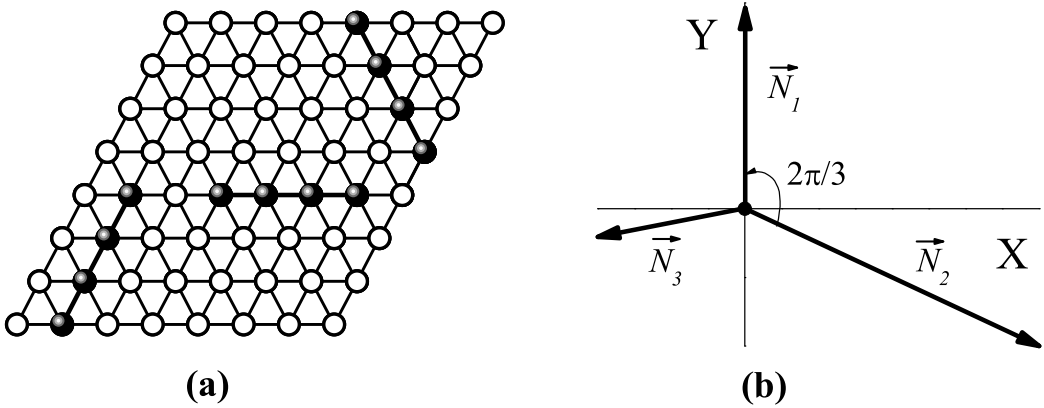}
\end{center} \caption{(a) Straight rigid rods adsorbed on triangular lattices. Solid circles (joined by thick lines) and empty circles represent tetramers ($k=4$) and empty sites, respectively. (b) Schematic representation of the set of vectors {$\vec{N}_{1}$ , $\vec{N}_{2}$ , $\vec{N}_{3}$ } for a triangular lattice}
\end{figure*}

Let $N_1$, $N_2$ and $N_3$ be the number of rods oriented along
directions 1, 2 and 3 on the surface, respectively. The total
number of $k$-mers is $N=N_1+N_2+N_3$. According to DiMarzio's
lattice theory (DiMarzio 1961), the number of ways
$\Omega(N_0,N_1,N_2,N_3)$ to pack the $N$ molecules such that
$N_i$ of them lie in the direction $i$ and there are $N_0$ empty
sites on the surface is given by
\begin{equation}
\begin{split}
\label{DiMarzio} \Omega(M,N_1,N_2,N_3) = & \frac{\prod_{j=1}^{3}
\left[M- \left( k-1 \right)N_j \right]!} {\left( N_0 \right)!
\prod_{i=1}^{3} \left( N_i \right)! \left( M! \right)^2 },
\end{split}
\end{equation}
where $N_0=M-k \sum_{i}^3N_{i}$.

Since different $k$-mers do not interact with each other, all
configurations of $N$ $k$-mers on $M$ sites are equally probable;
henceforth, the canonical partition function $Q(M,N_1,N_2,N_3,T)$
equals the total number of configurations,
$\Omega(M,N_1,N_2,N_3)$, times a Boltzmman factor including the
total interaction energy between $k$-mers and lattice sites, $k
\epsilon_0 N$
\begin{equation}
\label{particion}
Q(M,N_1,N_2,N_3,T)  =  q(T)^N \Omega(M,N_1,N_2,N_3) \exp \left(-{\beta k \epsilon_o N}\right),
\end{equation}
where $q(T)$ is the partition function for a single adsorbed
molecule, $\beta=1/k_BT$ (being $k_B$ the Boltzmann constant and
$T$ the temperature) and $\epsilon_o$ is the interaction energy
between every unit forming a $k$-mer and the substrate.

In the canonical ensemble the Helmholtz free energy
$F(M,N_1,N_2,N_3,T)$ relates to $\Omega(M,N_1,N_2,N_3)$ through
\begin{equation}
\begin{split}
\beta F(M,N_1,N_2,N_3,T) = & -\ln Q(M,N_1,N_2,N_3,T) \\
= & -N\ln q -\ln \Omega(M,N_1,N_2,N_3) + \beta k \epsilon_o N. \label{elf}
\end{split}
\end{equation}
Then, the remaining thermodynamic functions can be obtained from
the general differential form \cite{Hill}
\begin{equation}
\label{dF} dF = -SdT - \Pi dM + \mu dN,
\end{equation}
where $S$, $ \Pi $ and $ \mu $ designate the entropy, spreading
pressure and chemical potential respectively which by definition
are,
\begin{equation}
\label{FuncionesT} S = - \left({\partial F \over \partial
T}\right)_{M,N}  ;  \ \ \ \ \ \ \ \ \Pi = - \left({\partial F
\over \partial M}\right)_{T,N};  \ \ \ \ \ \ \ \ \ \mu =
\left({\partial F \over \partial N}\right)_{T,M}.
\end{equation}

\subsection{Isotropic distribution of adsorbed $k$-mers}

For the case of an isotropic distribution of the $k$-mers, i.e.,
$N_1=N_2=N_3=N/3$, \ref{DiMarzio} reduces to,
\begin{equation}
\label{GD_ISO} \Omega=\frac{\left\{ \left[ N_0+  \left( 2kN/3
\right) + \left( N/3  \right)  \right]!  \right\}^3} {N_0!
\left[\left( N/3 \right)! \right]^3 \left( M! \right)^{2}}.
\end{equation}
Applying the Stirling's approximation to \ref{GD_ISO} and
replacing in \ref{elf}, the Helmholtz free energy per site
$f=F/M$ can be written in terms of the intensive variables
$\theta$ and $T$,
\begin{equation}
\begin{split}
\label{FISO}
\beta f(\theta) =&-\left[3-\frac{(k-1)}{k}\theta\right]\ln\left[3-\frac{(k-1)}{k}\theta\right] 
+\frac{\theta}{k}\ln{\frac{\theta}{k}} +(1-\theta)\ln(1-\theta) \\
 & -(\theta-3)\ln{3}+\beta\epsilon_{o} \theta-\frac{\theta}{k}\ln
q(T).
\end{split}
\end{equation}
Then, the chemical potential and the entropy per site $s=S/M$
result
\begin{equation}
\begin{split}
\label{MUIS}
\beta \mu  = &(k-1)\ln\left[1-\frac{(k-1)}{3k}\theta\right]+\ln\left(\frac{\theta}{3k}\right) 
 -k\ln(1-\theta)-\ln K_e(T),
\end{split}
\end{equation}
and
\begin{equation}
\begin{split}
\label{SIS}
\frac{s(\theta)}{k_B}=&\left[3-\frac{(k-1)}{k}\theta\right]\ln\left[3-\frac{(k-1)}{k}\theta\right]-\frac{\theta}{k}\ln{\frac{\theta}{k}} -(1-\theta)\ln(1-\theta)+(\theta-3)\ln{3} \\
 & + \frac{\theta}{k} \left[ \ln q(T) + T \frac{d \ln q(T)}{d T}
\right],
\end{split}
\end{equation}
where $K_e(T)=q(T) \exp {(-\beta k \epsilon_o})$ is the
equilibrium constant.
\subsection{Anisotropic distribution of adsorbed \emph{k}-mers}
To introduce the effect of the orientational order in the GD
theory, it is convenient to rewrite the configurational factor in
\ref{DiMarzio} in terms of the nematic order parameter
$\vec{\delta}$ \cite{Wu},
\begin{equation}
\label{deltavector}
\vec{\delta}=\frac{\sum_{i=1}^{m}\vec{N}_{i}}{\sum_{i=1}^{m}\vert\vec{N}_{i}\vert}.
\end{equation}
$\vec{\delta}$ represents a general order parameter measuring the
orientation of the $k$-mers on a lattice with $m$ directions and
the set of vectors $\{\vec{N}_1,\vec{N}_2,\cdots,\vec{N}_m\}$ is
characterized by the following properties: $(i)$ each vector is
associated to one of the $m$ possible orientations (or directions)
for a $k$-mer on the lattice; $(ii)$ the $\vec{N}_i$'s lie in a
two-dimensional space (or are co-planar) and point radially
outward from a given point $P$ which is defined as coordinate
origin; $(iii)$ the angle between two consecutive vectors,
$\vec{N}_i$ and $\vec{N}_{i+1}$, is equal to $2\pi/m$; and $(iv)$
the magnitude of $\vec{N}_i$ is equal to the number of $k$-mers
aligned along the $i$-direction. Note that the $\vec{N}_i$'s have
the same directions as the $q$ vectors in \cite{Wu}. These
directions are not coincident with the allowed directions for the
$k$-mers on the real lattice.

In the case of a triangular lattice, as studied here, $m=3$, the
angle between $\vec{N}_i$ and $\vec{N}_{i+1}$ is $2\pi/3$ and \ref{deltavector} reduces to [see Figure 1(b)]:
\begin{equation}
\label{deltavectorm3}
\vec{\delta}=\frac{\vec{N}_{1}+\vec{N}_{2}+\vec{N}_{3}}{N_1+N_2+N_3}=
\frac{\vec{N}_{1}+\vec{N}_{2}+\vec{N}_{3}}{N},
\end{equation}
where $\vert\vec{N}_{i}\vert=N_{i}$ has been used for notational
convenience.

$\vec{\delta}$ can be expressed in Cartesian form as
$\vec{\delta}=\delta_{x} \hat{x} +\delta_{y} \hat{y}$, where
\begin{equation}
\delta_{x}=\frac{ N_{1}-\frac{1}{2} N_{2}-\frac{1}{2}N_{3}}{N},
\end{equation}
and
\begin{equation}
\delta_{y}=\frac{ \frac{\sqrt{3}}{2} N_{2}-\frac{\sqrt{3}}{2}
N_{3}}{N}.
\end{equation}
In addition,
\begin{equation}
\label{proyeciones}
\begin{split}
\theta=&\frac{k N}{M}=\frac{k \left(N_{1}+N_{2}+N_{3} \right)}{M}.
\end{split}
\end{equation}
Then, $N_{1}$, $N_{2}$ and $N_{3}$ can be written as a function of
$\delta_x$, $\delta_y$ and $\theta$,
\begin{equation}
\label{Ns}
\begin{split}
\frac{N_{1}}{M}=&\frac{\theta}{3k}\,\left(1+2\,\delta_{x}\right),\\
\frac{N_{2}}{M}=&\frac{\theta}{3k}\,\left(1-\delta_{x}+\sqrt{3}\,\delta_{y}\right),\\
\frac{N_{3}}{M}=&\frac{\theta}{3k}\,\left(1-\delta_{x}-\sqrt{3}\,\delta_{y}\right).
\end{split}
\end{equation}
Now, replacing \ref{Ns} in the DiMarzio configurational
factor \ref{DiMarzio} and using \ref{elf}, the
Helmholtz free energy per site can be written as,
\begin{equation}
\label{Fdelta}
\begin{split}
\beta f(\theta,\delta_x,\delta_y)=&(1-\theta)\ln(1-\theta)\\
&-\left[1-\frac{(k-1)}{3k} \left(1+2\,\delta_{x}\right) \theta \right]\ln\left[1-\frac{(k-1)}{3k} \left(1+2\,\delta_{x}\right) \theta \right]\\
&-\left[1-\frac{(k-1)}{3k}\left(1-\delta_{x}+\sqrt{3}\,\delta_{y}\right)\theta \right]\ln\left[1-\frac{(k-1)}{3k}\left(1-\delta_{x}+\sqrt{3}\,\delta_{y}\right)\theta \right]\\
&-\left[1-\frac{(k-1)}{3k}\left(1-\delta_{x}-\sqrt{3}\,\delta_{y}\right)\theta \right]\ln\left[1-\frac{(k-1)}{3k}\left(1-\delta_{x}-\sqrt{3}\,\delta_{y}\right)\theta \right]\\
&+\frac{\left(1+2\,\delta_{x}\right)}{3k}\theta\ln\left[\frac{\left(1+2\,\delta_{x}\right)}{3k}\theta\right]\\
&+\frac{\left(1-\delta_{x}+\sqrt{3}\,\delta_{y}\right)}{3k}\theta\ln\left[\frac{\left(1-\delta_{x}+\sqrt{3}\,\delta_{y}\right)}{3k}\right]\\
&+\frac{\left(1-\delta_{x}-\sqrt{3}\,\delta_{y}\right)}{3k}\theta\ln\left[\frac{\left(1-\delta_{x}-\sqrt{3}\,\delta_{y}\right)}{3k}\right]+\beta \epsilon_{o} \theta-\frac{\theta}{k}\ln q(T).\\
\end{split}
\end{equation}
Finally, from \ref{FuncionesT},
\begin{equation}
\label{MU_delta}
\begin{split}
\beta \mu(\theta,\delta_x,\delta_y) =& \frac{(k-1)}{3}\left[1+2\left(\frac{\partial \delta_x}{\partial \theta}\theta+\delta_x\right)\right]\ln\left[1-\frac{(k-1)}{3k} \left(1+2\,\delta_{x}\right) \theta \right]\\
&+\frac{(k-1)}{3}\left[1+\left(\sqrt{3}\frac{\partial\delta_y}{\partial\theta}-\frac{\partial\delta_x}{\partial\theta}\right)\theta-\delta_x+\sqrt{3}\delta_y\right]\ln\left[1-\frac{(k-1)}{3k}\left(1-\delta_{x}+\sqrt{3}\,\delta_{y}\right)\theta \right]\\
&+\frac{(k-1)}{3}\left[1-\left(\sqrt{3}\frac{\partial\delta_y}{\partial\theta}+\frac{\partial\delta_x}{\partial\theta}\right)\theta-\delta_x-\sqrt{3}\delta_y\right]\ln\left[1-\frac{(k-1)}{3k}\left(1-\delta_{x}-\sqrt{3}\,\delta_{y}\right)\theta \right]\\
&+\frac{1}{3}\left[1+2\left(\frac{\partial \delta_x}{\partial \theta}\theta+\delta_x\right)\right]\ln\left[\frac{\left(1+2\,\delta_{x}\right)}{3k}\theta\right]\\
&+\frac{1}{3}\left[1+\left(\sqrt{3}\frac{\partial\delta_y}{\partial\theta}-\frac{\partial\delta_x}{\partial\theta}\right)\theta-\delta_x+\sqrt{3}\delta_y\right]\ln\left[\frac{\left(1-\delta_{x}+\sqrt{3}\,\delta_{y}\right)}{3k}\theta\right]\\
&+\frac{1}{3}\left[1-\left(\sqrt{3}\frac{\partial\delta_y}{\partial\theta}+\frac{\partial\delta_x}{\partial\theta}\right)\theta-\delta_x-\sqrt{3}\delta_y\right]\ln\left[\frac{\left(1-\delta_{x}-\sqrt{3}\,\delta_{y}\right)}{3k}\theta\right]\\
&-k\ln(1-\theta)-\ln K_e(T).
\end{split}
\end{equation}
and
\begin{equation}
\label{S_delta}
\begin{split}
\frac{s(\theta,\delta_x,\delta_y)}{k_B}=&\left[1-\frac{(k-1)}{3k} \left(1+2\,\delta_{x}\right) \theta \right]\ln\left[1-\frac{(k-1)}{3k} \left(1+2\,\delta_{x}\right) \theta \right]\\
&+\left[1-\frac{(k-1)}{3k}\left(1-\delta_{x}+\sqrt{3}\,\delta_{y}\right)\theta \right]\ln\left[1-\frac{(k-1)}{3k}\left(1-\delta_{x}+\sqrt{3}\,\delta_{y}\right)\theta \right]\\
&+\left[1-\frac{(k-1)}{3k}\left(1-\delta_{x}-\sqrt{3}\,\delta_{y}\right)\theta \right]\ln\left[1-\frac{(k-1)}{3k}\left(1-\delta_{x}-\sqrt{3}\,\delta_{y}\right)\theta \right]\\
&-\frac{\left(1+2\,\delta_{x}\right)}{3k}\theta\ln\left[\frac{\left(1+2\,\delta_{x}\right)}{3k}\theta\right]\\
&-\frac{\left(1-\delta_{x}+\sqrt{3}\,\delta_{y}\right)}{3k}\theta\ln\left[\frac{\left(1-\delta_{x}+\sqrt{3}\,\delta_{y}\right)}{3k}\theta\right]\\
&-\frac{\left(1-\delta_{x}-\sqrt{3}\,\delta_{y}\right)}{3k}\theta\ln\left[\frac{\left(1-\delta_{x}-\sqrt{3}\,\delta_{y}\right)}{3k}\theta\right]\\
&-(1-\theta)\ln(1-\theta)+ \frac{\theta}{k} \left[ \ln q(T) + T
\frac{d \ln q(T)}{d T} \right],
\end{split}
\end{equation}
It is easy to see that, as $\vert\vec{\delta}\vert=0$, i.e.
$\delta_x = 0$ and $\delta_y = 0$, the isotropic case is recovered
and, consequently, \ref{Fdelta,MU_delta} reduce to
\ref{FISO}-\ref{MUIS}. In general, the calculation of the
adsorption isotherm and the configurational entropy of the adlayer
requires the knowledge of an analytical expression for the
dependence of the nematic order parameter on the coverage. For
this purpose, a free-energy-minimization approach can be applied
\cite{LANG14}. The procedure is as follows:

\begin{itemize}
\item[(1)] We choose $\delta_{y}=0$ and $\delta_{x}\neq 0$, this
leaves $N_2$ and $N_3$ in an isotropic state, see \ref{Ns}.
We can do this without losing any generality, since a pure nematic
state is given by molecules aligned in one direction only \cite{Oswald}.

\item[(2)] By differentiating \ref{Fdelta} (with
$\delta_{y}=0$) with respect to $\delta_x$ and setting the result
equal to zero, the function $\delta(\theta)$ is obtained.

\item[(3)] $\delta(\theta)$ is introduced in \ref{S_delta,MU_delta} and thus the adsorption isotherm and the
configurational entropy of the adlayer are obtained (without
orientational restrictions).
\end{itemize}
The points (2) and (3) can be easily solved through a standard
computing procedure; in our case, we used Maple software.

\section{3. Monte Carlo simulation}

In order to test the theory, an efficient hyper-parallel tempering
Monte Carlo (HPTMC) simulation method \cite{Yan,Hukushima} has been used. The HPTMC method
consists in generating a compound system of $R$ noninteracting
replicas of the system under study. The $i$-th replica is
associated with a chemical potential $\mu_{i}$. To determine the
set of chemical potentials, $\{\mu_i \}$, the lowest chemical
potential, $\mu_1$, is set in the isotropic phase where relaxation
(correlation) time is expected to be very short and there exists
only one minimum in the free energy space. On the other hand, the
highest chemical potential, $\mu_R$, is set in the nematic phase
whose properties we are interested in. Finally, the difference
between two consecutive chemical potentials, $\mu_i$ and
$\mu_{i+1}$ with $\mu_i < \mu_{i+1}$, is set as $\Delta \mu
=\left(\mu_{1} - \mu_R \right)/(R-1)$ (equally spaced chemical
potentials). The parameters used in the present study were as
follows: $R=25$, $\mu_1=-10$ and $\mu_R=10$. With these values of
the chemical potential, the corresponding values of the surface
coverage varied from $\theta_1(\mu_1) \approx 2 \times 10^{-4} $
to $\theta_R(\mu_R)\approx 0.99$ for $k=2$, and from
$\theta_1(\mu_1)\approx 2 \times 10^{-3}$ to
$\theta_R(\mu_R)\approx 0.96$ for $k=10$.

Under these conditions, the algorithm to carry out the simulation
process is built on the basis of two major subroutines: {\it
replica-update} and {\it replica-exchange}.

\noindent {\it Replica-update}: The adsorption-desorption
procedure is as follows: (1) One out of $R$ replicas is randomly
selected. (2) A linear $k$-uple of nearest-neighbor sites,
belonging to the replica selected in (1), is chosen at random.
Then, if the $k$ sites are empty, an attempt is made to deposit a
rod with probability $W = \min \left[1, \exp(\beta \mu)\right]$;
if the $k$ sites are occupied by units belonging to the same
$k$-mer, an attempt is made to desorb this $k$-mer with
probability $W = \min \left[1, \exp(-\beta \mu)\right]$; and
otherwise, the attempt is rejected. In addition, the displacement
(diffusional relaxation) of adparticles to nearest-neighbor
positions, by either jumps along the $k$-mer axis or reptation by
rotation around the $k$-mer end, must be allowed in order to reach
equilibrium in a reasonable time.

\noindent {\it Replica-exchange}: Exchange of two configurations
$X_i$ and $X_j$, corresponding to the $i$-th and $j$-th replicas,
respectively, is tried and accepted with probability $W=\min
\left[ 1,\exp{(-\Delta)}\right]$. Where $\Delta$ in a nonthermal
grand canonical ensemble is given by $[-\beta(\mu_{j}-\mu_{i})
\,(N_{j}-N_{i})]$, and $N_{i}$ ($N_{j}$) represents the number of
particles of the $i$-th ($j$-th) replica.

The complete simulation procedure is the following: (1)
replica-update, (2) replica-exchange, and (3) repeat from step (1)
$RM$ times. This is the elementary step in the simulation process
or Monte Carlo step (MCs).

For each value of the chemical potential $\mu_i$, the equilibrium
state can be well reproduced after discarding the first $r_0$ MCs.
Then, a set of $r$ samples in thermal equilibrium is generated.
The corresponding surface coverage $\theta_i(\mu_i)$ is obtained
through simple averages over the $r$ samples ($r$ MCs).
\begin{equation}
\theta_i(\mu_i)=\frac{1}{r} \sum_{t=1}^{r} \theta
\left[X_i(t)\right].
\end{equation}
In the last equation, $X_i$ stands for the state of the $i$-th
replica (at chemical potential $\mu_i$).

The configurational entropy $S$ of the adsorbate cannot be
directly computed. To calculate entropy, various methods have been
developed \cite{LANG6}. Among them, the thermodynamic
integration method is one of the most widely used and practically
applicable. The method in the grand canonical ensemble relies upon
integration of the chemical potential $\mu$ on coverage along a
reversible path between an arbitrary reference state and the
desired state of the system. This calculation also requires the
knowledge of the total energy $U$ for each obtained coverage.
Thus, for a system made of $N$ particles on $M$ lattice sites,
\begin{equation}
S(N,M,T)  =  S(N_0,M,T)+{U(N,M,T)-U(N_0,M,T) \over T} -{1\over T} \int_{N_0}^N{ \mu dN'}.
\end{equation}
In the present case $U(N,M,T)=0$ and the determination of the
reference state, $S(N_0,M,T)$, is trivial because $S(N_0,M,T)=0$
for $N_0 = 0$. Then, using intensive variables,
\begin{equation}
{s(\theta,T) \over k_B}= - {1 \over k_B T} \int_{0}^{\theta}{
\frac{\mu}{k} ~ d \theta'}. \label{ent}
\end{equation}

\section{4. Results}

In this section, the main characteristics of the thermodynamic
functions given in \ref{S_delta,MU_delta} will
be analyzed in comparison with simulation results and the main
theoretical models developed to treat the $k$-mers adsorption
problem. Three theories have been considered: the first is the
well-known FH approximation for straight rigid rods \cite{Flory,Huggins}; the
second is the GD approach for an isotropic distribution of
admolecules \cite{Guggenheim,DiMarzio}; and the third is the
recently developed SE model for the adsorption of polyatomics
\cite{LANG11,IJMP}.

The equations of the GD adsorption isotherm and the GD configurational entropy for an isotropic distribution of adsorbed rods were given in \ref{MUIS,SIS}, respectively. The corresponding expressions in the FH and SE
theories are as follows:
\begin{equation}
\beta  \mu  =  \ln \left(\frac{\theta}{k} \right) -k \ln  \left( 1-\theta \right) - \ln \left( \frac{c}{2} \right) 
-\ln K_e(T)\   \  \  \  \ \  \  \  \ \  \  \  \ \  \  \ \ (k
\geq 2) \ \ \  \  \  \ {\rm (FH)} \label{mufh},
\end{equation}
\begin{equation}
\begin{split}
\beta  \mu   = & \ln \left( \frac{\theta}{k} \right) -k \ln
\left( 1-\theta \right)
-\ln \left( \frac{c}{2} \right) +(1-\theta)(k-1) \ln \left[1- \frac{(k-1)}{k} \frac{2\theta}{c} \right]  \\
  & + \theta (k-1) \ln \left[1- \frac{(k-1)\theta}{k} \right] -\ln K_e(T) \   \  \  \  \ \  \  \  \ \  \  \  \ \  \ \  \  \
\  \  \  \ \  \  \  \  \ \  \  \ \ \ {\rm (SE)}, \label{isoemp}
\end{split}
\end{equation}
\begin{equation}
{s(\theta) \over k_B}  =   - {\theta \over k}\ln{\theta \over k}
-\left(1-\theta \right)\ln \left(1-\theta \right)  - {\theta \over k} \left[k-1- \ln \left( c \over 2 \right) \right]  + \frac{\theta}{k} \left[ \ln q(T) + T \frac{d \ln q(T)}{d T}
\right] \ \ \  \ \  \  \ \ \ {\rm (FH) \label{fh}},
\end{equation}
and
\begin{equation}
\begin{split}
{s(\theta) \over k_B}  = &  - {\theta \over k}\ln{\theta \over k}
-\left(1-\theta \right)\ln \left(1-\theta \right) + \theta \left[{1 \over 2} - {c \over 4} + {1 \over k} \ln
\left( c \over 2 \right) \right] \\
 & +{1 \over 2} {k \over \left(k-1 \right)} \left[1-{\left(k-1
\right)^2 \over k^2} \theta^2 \right]\ln \left[1-{\left(k-1
\right) \over k} \theta \right] \\
 & - {c \over 4} \left[\theta + {k \left(c-4 \right)+4 \over 2
\left(k-1 \right)} \right] \left[1-{2 \left(k-1 \right) \over c k}
\theta \right]\ln \left[1-{2 \left(k-1 \right) \over c k} \theta
\right] \\
 & + \frac{\theta}{k} \left[ \ln q(T) + T \frac{d \ln q(T)}{d T}
\right]\ \ \ \ \ \ \  {\rm (SE) \label{se}}.
\end{split}
\end{equation}

The computational simulations have been developed for triangular
$L \times L$ lattices with $L/k = 20$ and periodic boundary
conditions. With this size of the lattice we verified that finite
size effects are negligible. As mentioned in Ref. [\cite{Ghosh}], the relaxation time increases very quickly as the $k$-mer
size increases. Consequently, MC simulations for large adsorbates
are very time consuming and may produce artifacts related to
non-accurate equilibrium states. In order to discard this
possibility, equilibration times $r_0$ of the order O($10^7$ MCs)
were used in this study.

An extensive comparison among the new adsorption isotherm [\ref{MU_delta}, solid line], the simulation data (symbols), and
the isotherm equations obtained from the analytical approaches
depicted as GD [\ref{MUIS}, dashed line], FH [\ref{mufh}, dashed and dotted line], and SE [\ref{isoemp},
dotted line] is shown in Figure 2: (a) $k = 3$, (b) $k = 8$ and (c)
$k = 10$. In the case of\ref{MU_delta}, $\delta(\theta)$
was obtained by following the minimization procedure described at
the end of Sec. II. In addition, $q(T)$ is set equal to one in the
theoretical equations (vibrational and rotational degrees of
freedom of the adsorbed molecules are not considered in the
simulations).

\begin{figure*}
\begin{center}
\includegraphics[width=0.45\textwidth]{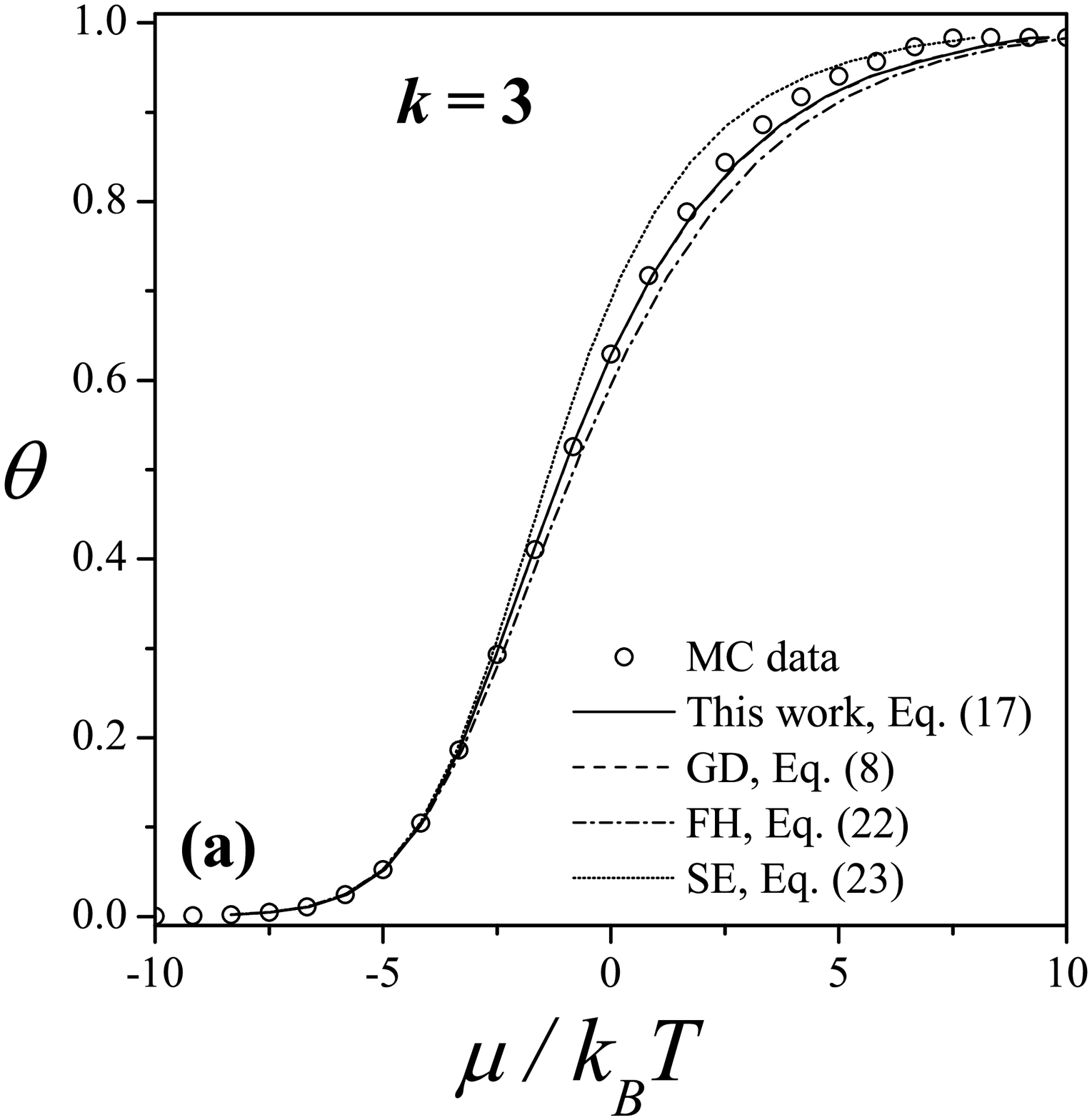}\hspace*{1.0cm}
\includegraphics[width=0.45\textwidth]{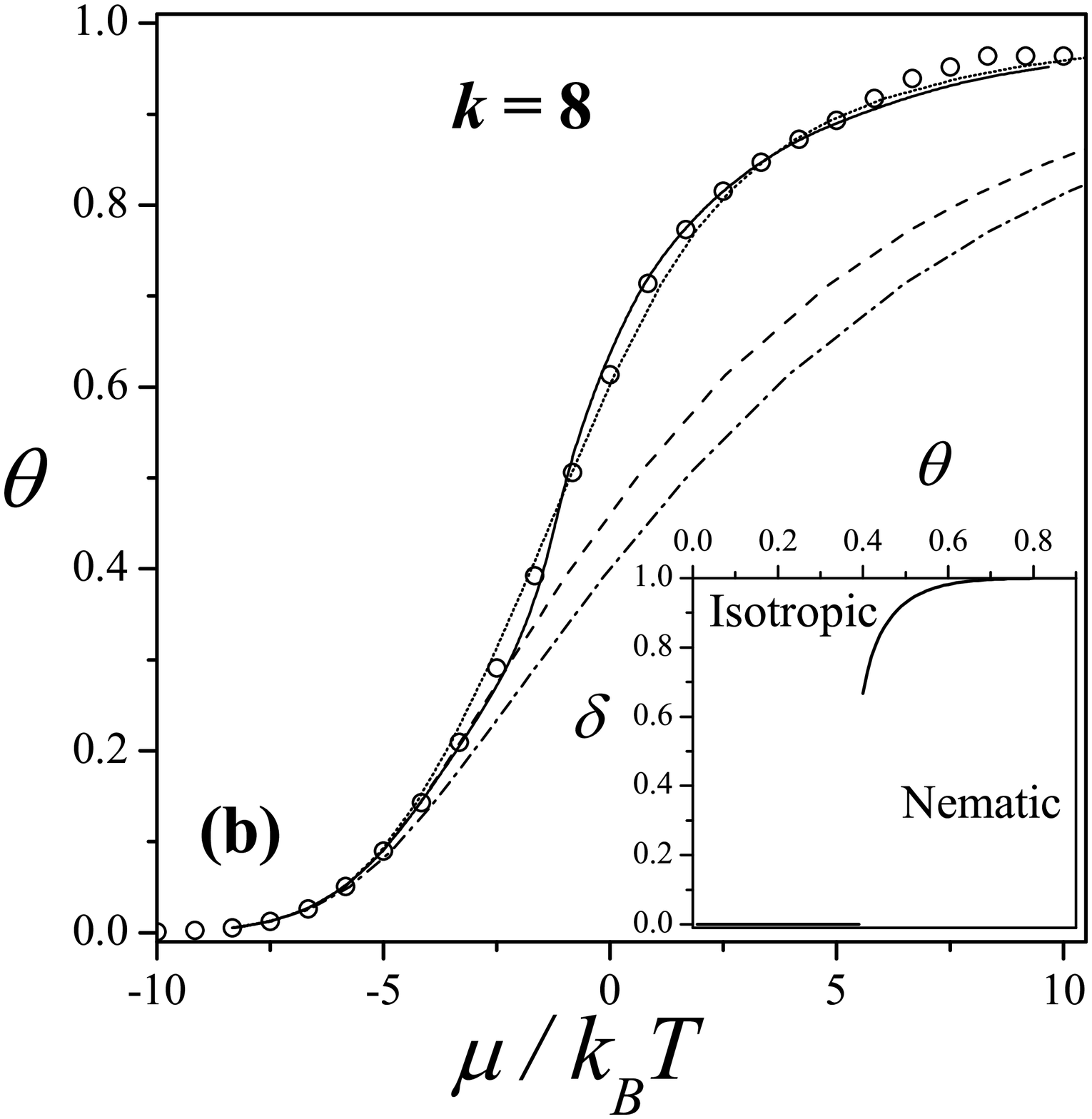}\\
\includegraphics[width=0.45\textwidth]{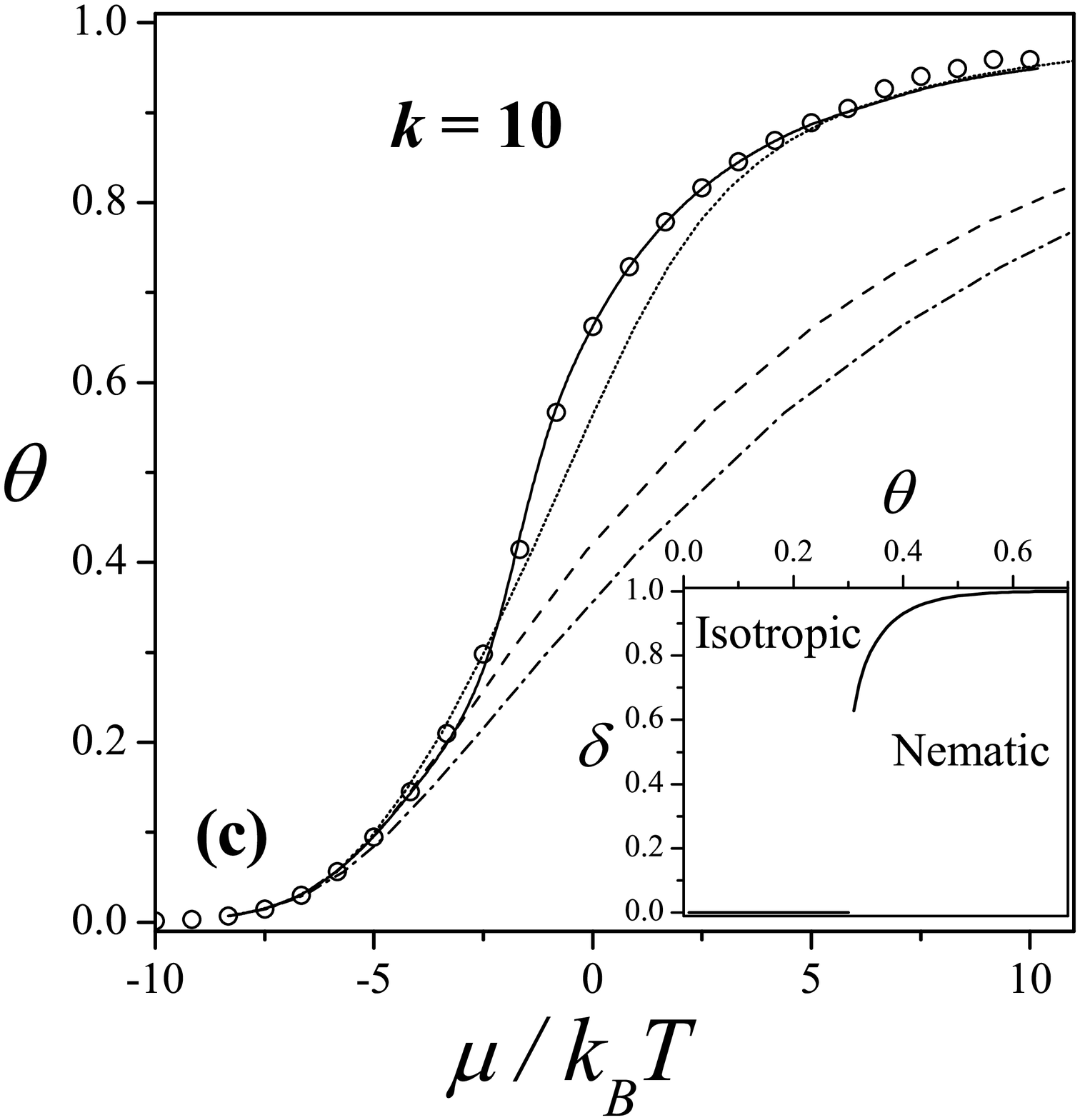}\hspace*{1.0cm}
\includegraphics[width=0.45\textwidth]{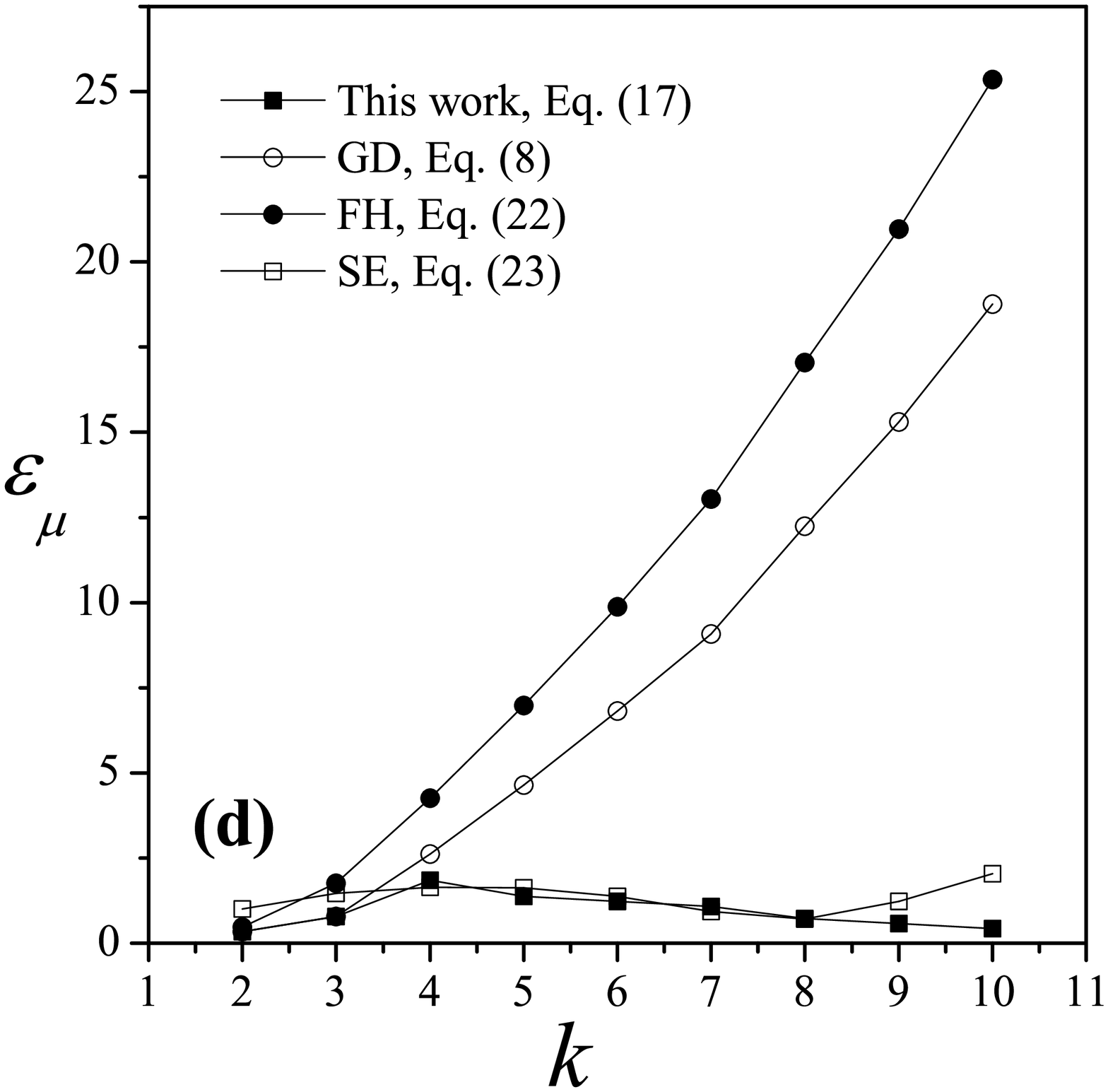}
\end{center}
\caption{Adsorption isotherms for rigid $k$-mers on a triangular
lattice: (a) $k = 3$, (b) $k = 8$, and (c) $k = 10$. Symbols
represent the MC results, and lines correspond to different
theoretical approaches as indicated in part (a). The corresponding
order parameters, obtained from the minimization of the free
energy in \ref{Fdelta}, are shown in the insets. (d) Average
percent error in the chemical potential, $\epsilon_{\mu}$, as a
function of $k$ for the different approximations studied in this
contribution}
\end{figure*}

In part (a), the behavior of the different approaches can be
explained as follows. The new theory and GD agree very well with
the simulation results for coverage values of up to $\theta
\approx 0.8$; however, the disagreement between theoretical and
simulation data increases for larger $\theta$ values. The
coincidence between the new theory and GD results is due to the
fact that, for small values of $k$ ($k < 4$), the function
$\delta(\theta)$ minimizing the free energy is $\delta(\theta) =
0$ and, under this condition, \ref{MU_delta} and \ref{MUIS} become identical. However, SE provides a good
approximation with very small differences between simulation and
theoretical results in all ranges of coverage.

Let us consider now the case of $k = 8$ [Figure 2(b)]. The agreement
between simulation and analytical data is very good for small
values of coverage. However, as the surface coverage is increased,
two different behaviors are observed. Although SE and the new
theory provide good results, the classical FH and GD
approximations fail to reproduce the simulation data. The
differences between GD and the theory in \ref{MU_delta} are
associated with the behavior of the order parameter
$\delta(\theta)$, which is shown in the inset of the figure. The
functionality of $\delta$ with coverage is indicative of the
existence of nematic order for $\theta > 0.4$. Even though this
result is not exact, the inclusion of $\delta(\theta)$ in \ref{MU_delta} leads to an extremely good approximation of the
adsorption isotherm.

The marked jump observed in the curve of the order parameter of a
function of the coverage [see inset of Figure 2(b)] is indicative of
the existence of a first-order phase transition in the adlayer.
This behavior differs from that obtained for square lattices
\cite{LANG14}, where the continuous variation of
the order parameter with density indicates clearly the presence of
a second-order phase transition in the adsorbed layer. This point
is extensively discussed in the recent paper by \cite{Dhar}.

Figure 2(c) is devoted to the analysis of large adsorbates ($k =
10$, in the case of the figure). The results are very clear: (1)
FH and GD predict a smaller $\theta$ than the simulation data over
the entire range of coverage; (2) SE agrees very well with the
simulation results for small and high values of the coverage;
however, the disagreement turns out to be significantly large in a
wide range of coverage ($0.3 < \theta < 0.9$); and (3) in the case
of the new isotherm, the results are excellent and represent a
significant advance with respect to the existing development of
$k$-mer thermodynamics.

In order to compute the accuracy of each theory, the differences
between theoretical and simulation data can be very easily
rationalized by using the average percent error in the chemical
potential $\varepsilon_{\mu}$, which is defined as,
\begin{equation}
\varepsilon_{\mu} = \frac{1}{N} \left(\sum_{\theta}
\left|\frac{\mu_{\rm sim}-\mu_{\rm appr}}{\mu_{\rm
sim}}\right|_{\theta} \right) \times 100 \%, \label{errormu}
\end{equation}
where $\mu_{\rm sim}$ ($\mu_{\rm appr}$) represents the value of
the chemical potential obtained by using the MC simulation
(analytical approach). Each pair of values ($\mu_{\rm
sim},\mu_{\rm appr}$) is obtained at fixed $\theta$. The sum runs
over the $N$ points of the simulation adsorption isotherm (in this
case, $N=25$ for all $k$).

The dependence of $\varepsilon_{\mu}$ on the $k$-mer size is shown
in Figure 2(d) for the different theoretical approximations. Several
conclusions can be drawn from the figure:

\begin{itemize}

\item[1)] In the FH and GD cases, $\varepsilon_{\mu}$ increases
monotonically with increasing $k$ and the disagreement between MC
and analytical data turns out to be very large (larger than 5$\%$)
for $k \geq 5$ and $k \geq 6$, respectively.

\item[2)] For the SE theory, there exists a range of $k$ ($2 \leq
k \leq 7$) where $\varepsilon_{\mu}$ remains almost constant
around 1.5$\%$ and SE provides a very good fitting of the
simulation data. However, for $k \geq 8$, the differences between
simulation and theoretical data increase with $k$. This deviation
is associated with the appearance of an I-N phase transition in
the adlayer for $k > 7$ \cite{JCP7}, which is
not covered by the SE theory.

\item[3)] The agreement between the equation reported here [\ref{MU_delta}] and the simulation data is excellent over the
whole coverage range. This result provides valuable insight into
how the adsorption process takes place. Namely, for $k \geq 7$ and
intermediate densities, it is more favorable for the rods to align
spontaneously because the resulting loss of orientational entropy
is compensated for by the gain of translational entropy.

\item[4)] The comparison with previous results obtained for square
lattices \cite{LANG14} reveals that, for a fixed
value of $k$, (i) $\varepsilon_{\mu}$ increases with the
connectivity for FH and GD theories, and (ii) in the case of SE
approach and the new \ref{MU_delta}, $\varepsilon_{\mu}$
does not change significantly as the lattice geometry is varied.

\end{itemize}

The differences between the approaches analyzed in this work can
be also appreciated by comparing the coverage dependence of the
configurational entropy per site, which is presented in Figure 3 for
the same cases studied in Figure 2. and triangular lattices,
respectively. The overall behavior of $s(\theta)$ can be
summarized as follows: for $\theta \rightarrow 0$ the entropy
tends to zero. For low coverage, $s(\theta)$ is an increasing
function of $\theta$, reaches a maximum at $\theta_m$, then
decreases monotonically for  $\theta > \theta_m$. The position of
$\theta_m$ shifts to higher coverage as the $k$-mer size is
increased. In the limit $\theta \rightarrow 1$ the entropy tends
to a finite value, which is associated with the different ways to
arrange the $k$-mers at full coverage. This value depends on $k$.

\begin{figure*}
\begin{center}
\includegraphics[width=0.45\textwidth]{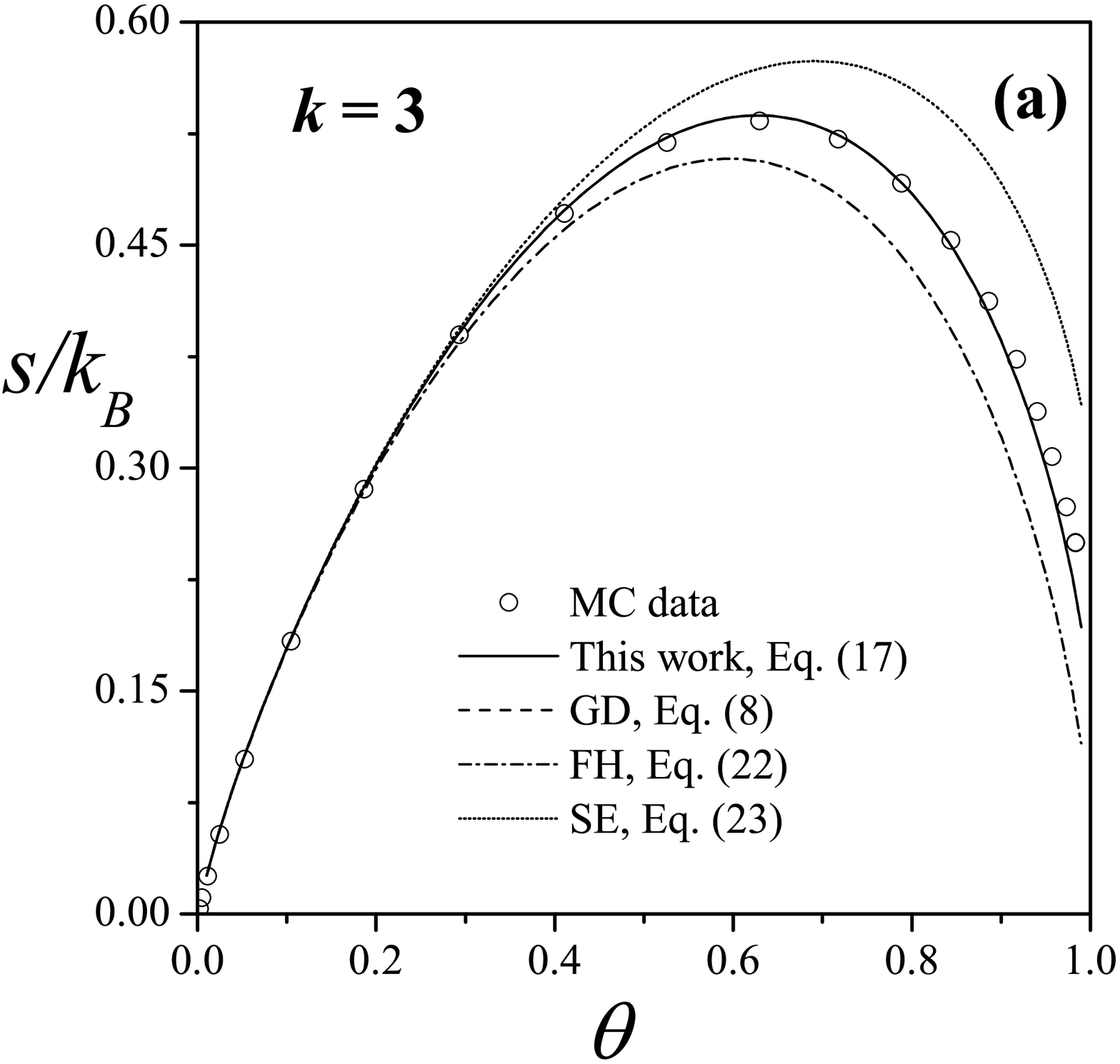}\hspace*{1.0cm}
\includegraphics[width=0.45\textwidth]{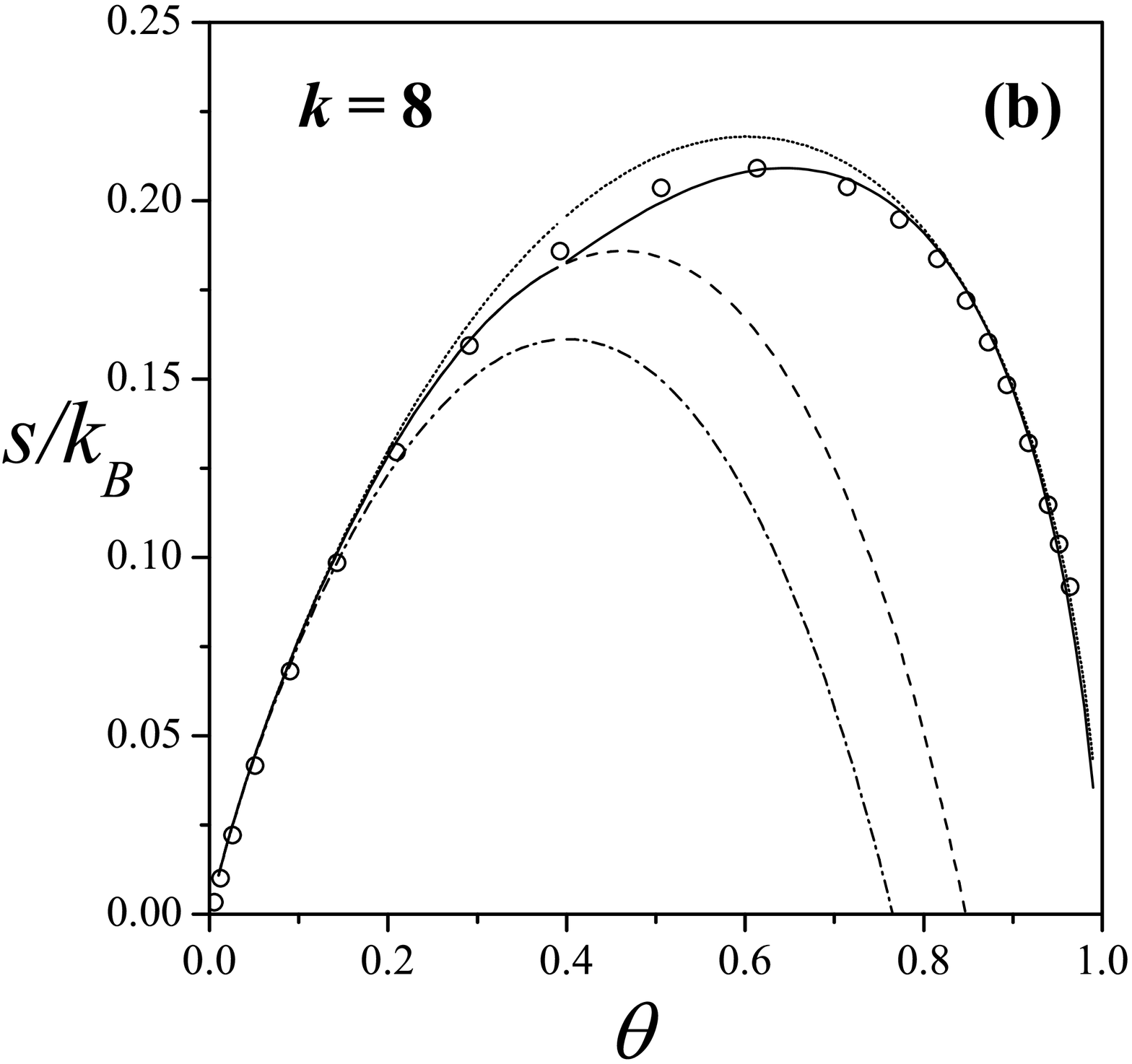}\\
\includegraphics[width=0.45\textwidth]{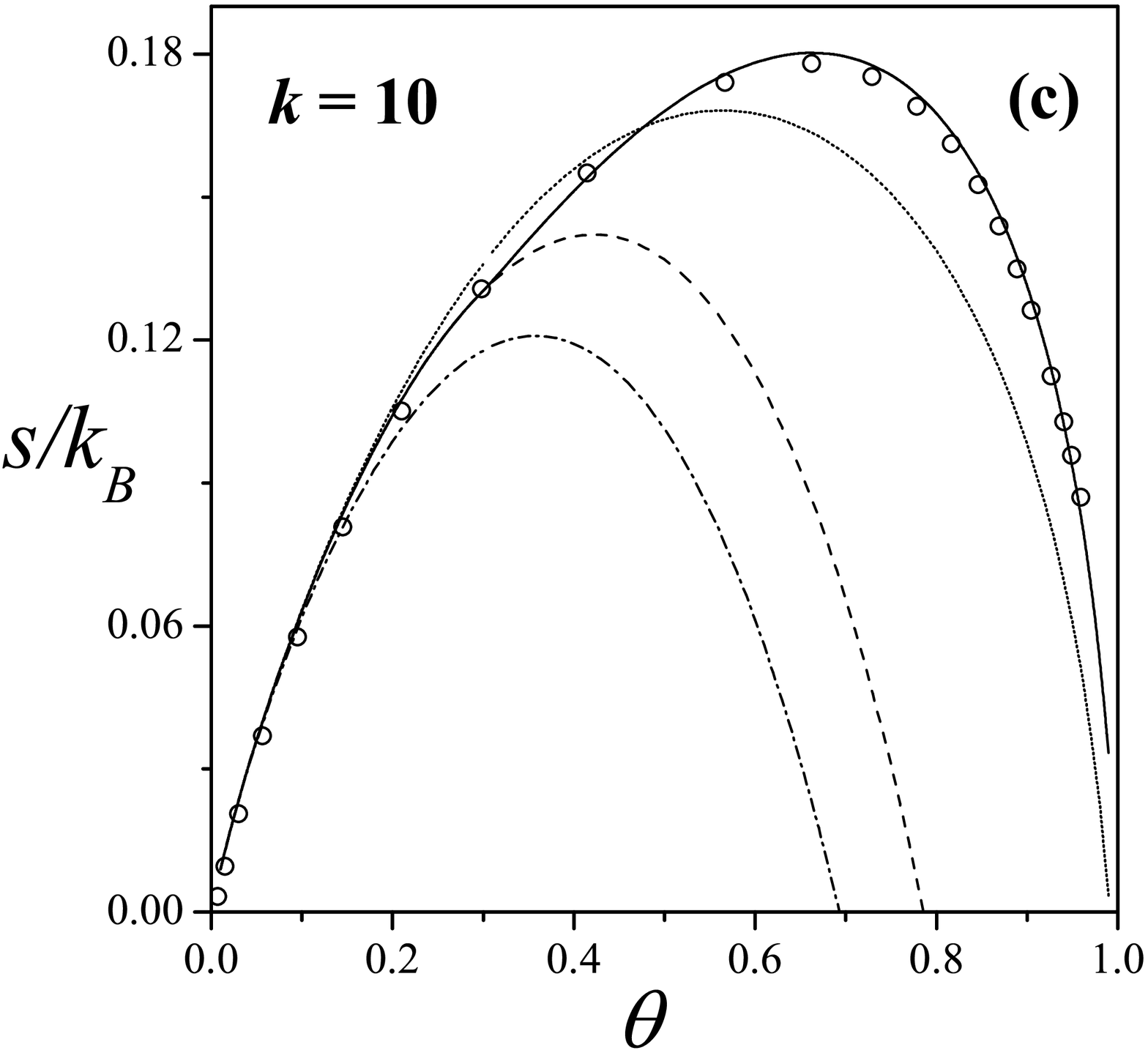}\hspace*{1.0cm}
\includegraphics[width=0.45\textwidth]{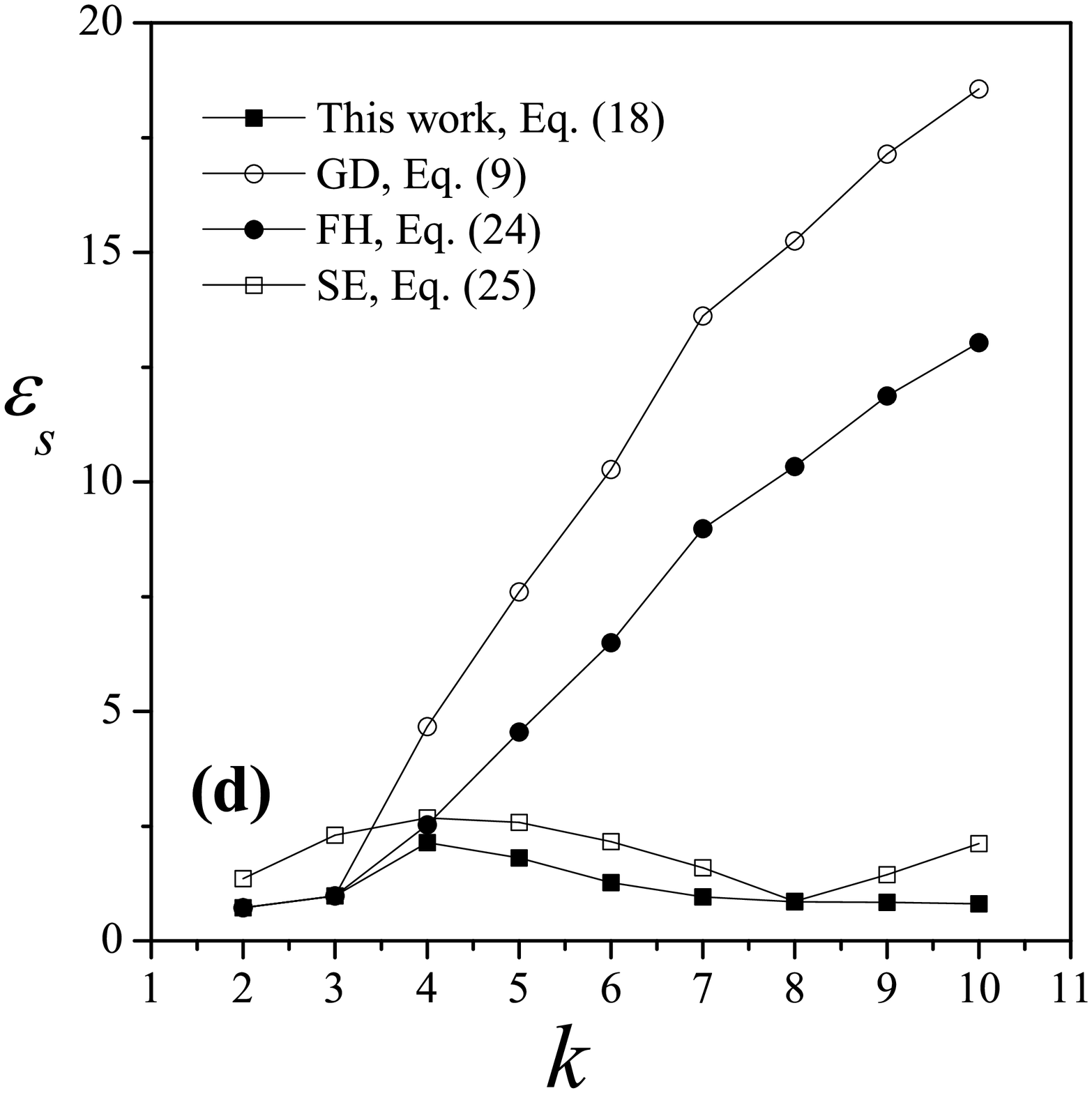}
\end{center}
\caption{Same as Figure 2 for the configurational entropy of the
adlayer}
\end{figure*}

As in Figure 2, GD and FH appear as good approximations in the
low-surface coverage region, but the disagreement turns out to be
significantly large for $s(\theta_m)$ and $s(\theta=1)$. On the
other hand, SE shows a good agreement with MC simulations up to
adsorbate sizes of $k \approx 8$. Finally, in the case of \ref{S_delta}, the agreement is notable for all $\theta$,
reproducing the MC results for $s(\theta_m)$ and $s(\theta=1)$.

As in the case of the chemical potential, an average percent error
($\varepsilon_{s}$) was calculated for the difference between
simulation and theoretical predictions. In this case,
\begin{equation}
\varepsilon_{s} = \frac{1}{N} \left(\sum_{\theta}
\left|\frac{s_{\rm sim}-s_{\rm appr}}{s_{\rm sim}}\right|_{\theta}
\right) \times 100 \%, \label{errors}
\end{equation}
where $s_{\rm sim}$ ($s_{\rm appr}$) represents the value of the
configurational entropy per site obtained by using the MC
simulation (analytical approach). As in \ref{errormu}, each
pair of values ($s_{\rm sim},s_{\rm appr}$) is obtained at fixed
$\theta$ and $N=25$.

The behavior of $\varepsilon_{s}$ is similar to that observed in
Figure 2(d). However, two main differences can be marked: (1) FH
performs better than GD for all values of $k$, and (2) the
differences between SE and \ref{errormu} are more notorious,
with \ref{errormu} being the most accurate for all cases.

Finally, analysis of experimental results have been carried out in
order to test the applicability of the model proposed here. For
this purpose, experimental adsorption isotherms of n-hexane in 5A
zeolites, previously compiled by Silva and Rodrigues \cite{Silva1}, were
analyzed in terms of \ref{MU_delta}. Given that the
experimental data were reported in adsorbed amount (g/100 g
adsorbed) as a function of pressure, the theoretical isotherms
were rewritten in terms of the pressure $p$ and the adsorbed
amount $Q$ as fitting quantities. Thus, assuming that the adsorbed
phase is in equilibrium with a ideal gas phase, the pressure $p$
can be written as $p \propto \exp{(\beta \mu})$. In addition,
$\theta=Q/Q_{max}$, where $Q_{max}$ represents the maximum
adsorbed amount. This choice allows us a direct comparison of \ref{MU_delta} with the results obtained in Ref. [\cite{Silva1}].

\begin{figure}
\begin{center}
\includegraphics[width=0.45\textwidth]{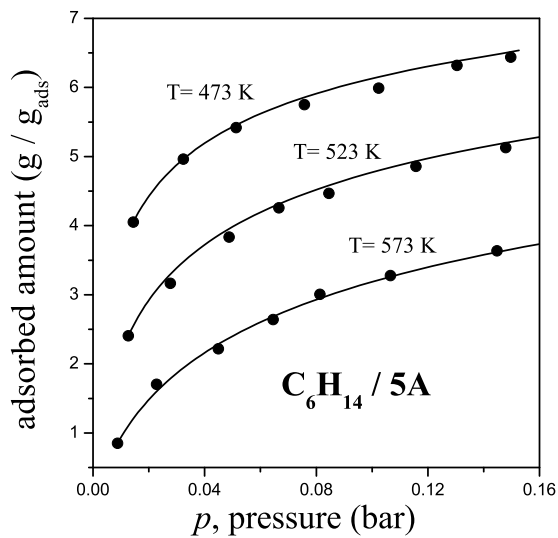}
\end{center}
\caption{Comparison between experimental and theoretical
adsorption isotherms (adsorbed amount $Q$ vs pressure $p$) for
$C_6 H_{14}$ adsorbed in 5A zeolite. Symbols represent
experimental data from Ref. [\cite{Silva2}] and lines correspond to
results from \ref{MU_delta}. The parameters used in the
fitting procedure are listed in Table I}
\end{figure}

As is common in the literature \cite{LANG10,IECR1}, a ``bead segment" chain model of the molecules was
adopted, in which each methyl (bead) group occupies one adsorption
site on the surface. Under this consideration, $k =6$ is set in
the fitting data corresponding to $C_6$. In this scheme, a set of
isotherms of n-hexane in 5A zeolites for different temperatures
were correlated by using only one value of $Q_{max}$ and a
temperature dependent $K_e(T)$ as adjustable parameters. The
results are presented in Figure 4 and the fitting parameters are
listed in Table I. A very good agreement between experimental and
theoretical data is observed. In addition, the value obtained for
the saturation adsorbed amount $Q_{max}=12.1$  is consistent with
previous results reported in Refs. [\cite{LANG10,Silva2}].

\begin{table}[b]
\begin{center}
\begin{tabular}{|c|c|c|}
\hline  \hline
{\rm Temperature} \ {\rm (K)} & $Q_{max}$ \ {\rm (g/100 $g_{ads}$}{\rm )} & $K_{e}$ \ {\rm (bar$^{-1}$ )}\\
\hline
473 & 12.1 & 0.109 \\
\hline
523 & 12.1 & 0.402 \\
\hline
573 & 12.1 & 1.597 \\
\hline

\end{tabular}
\caption{Table of parameters used in the fitting of Figure 4}
\end{center}
\end{table}

In summary, the analysis presented in Figures 2-4 demonstrates that
(1) explicitly considering the isotropic and nematic states
occurring in the adlayer at different densities is crucial to
understanding the adsorption process of rigid rods, and (2) \ref{Fdelta,MU_delta} provide a very good theoretical
framework and compact equations to consistently interpret
thermodynamic adsorption experiments of polyatomic species.

\section{5. Conclusions}

The adsorption process of straight rigid rods of length $k$ on
triangular lattices has been studied via grand canonical Monte
Carlo simulations, theory and analysis of experimental data. The
proposed theoretical formalism, based on a generalization of the
GD statistics, is capable of including the effects of the I-N
phase transition occurring at intermediate densities on the
thermodynamic functions of the system.

The results obtained (1) represent a significant qualitative
advance with respect to former developments on $k$-mer
thermodynamics; (2) demonstrates that explicitly considering the
isotropic and nematic states occurring in the adlayer at different
densities is crucial to understanding the adsorption process of
rigid rods; and (3) provide a very good theoretical framework and
compact equations to consistently interpret thermodynamic
adsorption experiments of polyatomic species.

\acknowledgement

This work was supported in part by CONICET (Argentina) under
project number PIP 112-200801-01332; Universidad Nacional de San
Luis (Argentina) under project 322000 and the National Agency of
Scientific and Technological Promotion (Argentina) under project
PICT-2010-1466.

\end{document}